\RequirePackage{lineno}
\documentclass[prl,twocolumn,showpacs,amsmath,amssymb]{revtex4-1}
\usepackage{overpic}
\usepackage{graphicx}
\usepackage{epsfig}
\usepackage{dcolumn}
\usepackage{bm}
\usepackage{rotating}
\usepackage{multirow}
\usepackage{subfigure}
\usepackage{hhline}
\usepackage{amssymb}
\usepackage{lineno}
\usepackage{hyperref}
\usepackage{mathrsfs}
\usepackage{verbatim}
\usepackage{amsmath}
\usepackage{amssymb}
\usepackage{amsfonts}
\usepackage{amsbsy}
\hypersetup{colorlinks,linkcolor=blue,anchorcolor=blue,citecolor=blue}
\include{def-com}

\let\oldequation\equation
\let\oldendequation\endequation

\renewenvironment{equation}
  {\linenomathNonumbers\oldequation}
  {\oldendequation\endlinenomath}

\newcommand{\mev}{\mathrm{MeV}}

\newcommand{\mevcc}{\mathrm{MeV}/c^2}

\begin{document}

\title{\boldmath Test of $C\!P$ Symmetry in Hyperon to Neutron Decays}

\author{
\begin{small}
\begin{center}
M.~Ablikim$^{1}$, M.~N.~Achasov$^{13,b}$, P.~Adlarson$^{75}$, X.~C.~Ai$^{81}$, R.~Aliberti$^{36}$, A.~Amoroso$^{74A,74C}$, M.~R.~An$^{40}$, Q.~An$^{71,58}$, Y.~Bai$^{57}$, O.~Bakina$^{37}$
% ...
\\
\vspace{0.2cm}
(BESIII Collaboration)\\
\vspace{0.2cm} {\it
$^{1}$ Institute of High Energy Physics, Beijing 100049, People's Republic of China\\
% ...
$^{13}$ G.I. Budker Institute of Nuclear Physics SB RAS (BINP), Novosibirsk 630090, Russia\\
% ...
$^{36}$ Johannes Gutenberg University of Mainz, Johann-Joachim-Becher-Weg 45, D-55099 Mainz, Germany\\
$^{37}$ Joint Institute for Nuclear Research, 141980 Dubna, Moscow region, Russia\\
% ...
$^{40}$ Liaoning Normal University, Dalian 116029, People's Republic of China\\
% ...
$^{57}$ Southeast University, Nanjing 211100, People's Republic of China\\
$^{58}$ State Key Laboratory of Particle Detection and Electronics, Beijing 100049, Hefei 230026, People's Republic of China\\
% ...
$^{71}$ University of Science and Technology of China, Hefei 230026, People's Republic of China\\
% ...
$^{74}$ University of Turin and INFN, (A)University of Turin, I-10125, Turin, Italy; (B)University of Eastern Piedmont, I-15121, Alessandria, Italy; (C)INFN, I-10125, Turin, Italy\\
$^{75}$ Uppsala University, Box 516, SE-75120 Uppsala, Sweden\\
% ...
$^{81}$ Zhengzhou University, Zhengzhou 450001, People's Republic of China\\
\vspace{0.2cm}
% ...
$^{b}$ Also at the Novosibirsk State University, Novosibirsk, 630090, Russia\\
% ...
}
\end{center}
\vspace{0.4cm}
\end{small}
}

\begin{abstract}
The quantum entangled $J/\psi \to
\Sigma^{+}\bar{\Sigma}^{-}$ pairs from
$(1.0087\pm0.0044)\times10^{10}$ $J/\psi$ events taken by the BESIII
detector are used to study the non-leptonic two-body weak decays
$\Sigma^{+} \to n \pi^{+}$ and $\bar{\Sigma}^{-} \to \bar{n}
\pi^{-}$. The $C\!P$-odd weak decay parameters of the decays
$\Sigma^{+} \to n \pi^{+}$ ($\alpha_{+}$) and $\bar{\Sigma}^{-} \to
\bar{n} \pi^{-}$ ($\bar{\alpha}_{-}$) are determined to be
$-0.0565\pm0.0047_{\rm stat}\pm0.0022_{\rm syst}$ and $0.0481\pm0.0031_{\rm
  stat}\pm0.0019_{\rm syst}$, respectively. The decay parameter
$\bar{\alpha}_{-}$ is measured for the first time, and the accuracy of
$\alpha_{+}$ is improved by a factor of four compared to the previous
results. The simultaneously determined decay parameters allow the
first precision $C\!P$ symmetry test for any hyperon decay with a
neutron in the final state with the measurement of
$A_{C\!P}=(\alpha_{+}+\bar{\alpha}_{-})/(\alpha_{+}-\bar{\alpha}_{-})=-0.080\pm0.052_{\rm
  stat}\pm0.028_{\rm syst}$. Assuming $C\!P$ conservation, the average
decay parameter is determined as $\left<
\alpha_{+}\right>=(\alpha_{+}- \bar{\alpha}_{-})/2 =
-0.0506\pm0.0026_{\rm stat}\pm0.0019_{\rm syst}$, while the ratios
$\alpha_{+}/\alpha_{0}$ and $\bar{\alpha}_{-}/\bar\alpha_{0}$ are $-0.0490\pm0.0032_{\rm stat}\pm0.0021_{\rm syst}$ and $-0.0571\pm0.0053_{\rm stat}\pm0.0032_{\rm syst}$, where $\alpha_{0}$ and
$\bar\alpha_{0}$ are the decay parameters of the decays $\Sigma^{+}
\to p \pi^{0}$ and $\bar{\Sigma}^{-} \to \bar{p} \pi^{0}$, respectively.
\end{abstract}

\maketitle
\oddsidemargin -0.2cm
\evensidemargin -0.2cm

Charge-parity ($C\!P$) violation is one of Sakharov's three essential
conditions for understanding the matter-antimatter asymmetry in the
universe~\cite{Sakharov:1967dj}.  Despite the established presence of
$C\!P$ violation in the decays of $K$, $B$, and $D$
mesons~\cite{Christenson:1964fg, Belle:2001qdd, BaBar:2001ags,
  Belle:2004nch, BaBar:2004gyj, LHCb:2019hro}, the standard model (SM)
of particle physics, as described by the Kobayashi-Maskawa mechanism,
is insufficient in fully explaining the preponderance of matter over
antimatter in the universe~\cite{Peskin:2002mm}.  As a result, it is
imperative to continue searching for new sources of $C\!P$
violation, particularly in the hyperon sector~\cite{Bigi:2017eni}.
The non-leptonic decays of spin-1/2 hyperons are suitable for $C\!P$
violation studies. In such decays, the decay asymmetry parameters
$\alpha$, $\beta$ and $\gamma$ are defined in terms of the
$S$-wave (parity violating) and $P$-wave (parity conserving)
amplitudes~\cite{Lee:1957qs}:

\begin{equation*}
\begin{split}
\alpha=\frac{2 Re (S^*P)}{|S|^{2}+|P|^{2}},\beta=\frac{2Im(S^*P)}{|S|^{2}+|P|^{2}},\gamma=\frac{|S|^{2}-|P|^{2}}{|S|^{2}+|P|^{2}}.
\label{eq:parameter}
\end{split}
\end{equation*}
Only two of these asymmetry parameters are independent. The magnitude of polarization of spin-1/2 hyperons can be inferred in two-body weak
decays due to their self-analyzing nature. The polar angle
distribution of the daughter nucleons is given by
$\mathrm{d}N/\mathrm{d}\Omega = \frac{1}{4\pi}( 1 + \alpha\mathbf{P}
\cdot \mathbf{\hat{p}})$.  Here, $\mathbf{P}$ is the hyperon
polarization vector, and $\mathbf{\hat{p}}$ is the unit vector along
the nucleon momentum in the hyperon rest frame. Correspondingly, the
decay asymmetry parameter of the anti-hyperon is denoted as
$\bar{\alpha}$.  Because $\alpha$ and $\bar{\alpha}$ are $C\!P$-odd,
$\it{A_{C\!P}}=(\alpha+\bar{\alpha})/(\alpha-\bar{\alpha})$ can be
used to test $C\!P$ conservation ~\cite{Okubo:1958zza,
  Pais:1959zza}. A non-zero value of $\it{A_{C\!P}}$ would indicate
$C\!P$ violation.

Theoretically, there are two predictions for $C\!P$ violation in
non-leptonic two-body weak decays of $\Sigma$.  In the seminal work by
Donoghue et al., the $C\!P$ violation contribution in
$\Sigma^{+} \rightarrow n \pi^{+}$ was predicted to be $-1.6 \times
10^{-4}$~\cite{Donoghue:1986hh}.  The most recent study by Tandean and
Valencia used heavy baryon Chiral perturbation theory and predicted
the $C\!P$ violation of $\Sigma^{+} \rightarrow n \pi^{+}$ to be
$3.9\times10^{-4}$.  Although the above two predictions are at the
same level, Ref.~\cite{Donoghue:1986hh} does not consider the $P$-wave
factorization contribution, which can change the prediction
by a factor of ten. To determine the SM $C\!P$ violation contribution,
the experimentally determined asymmetry parameters are used as part of
the input. Due to the large experimental uncertainty of the
$\Sigma^{+} \rightarrow n \pi^{+}$ asymmetry parameter $\alpha_+$, the
uncertainties in the $C\!P$ violation estimations of $\Sigma^{+}
\rightarrow n \pi^{+}$ are greater than those of other hyperons, and
the predicted $C\!P$ violation is an order of magnitude greater than
those of $\Sigma^{+} \rightarrow p \pi^{0}$, $\Sigma^{-} \rightarrow n
\pi^{-}$ and $\Lambda \rightarrow p \pi^{-}$~\cite{Tandean:2002vy}.

Recently it was pointed out that the experimental value of the decay
asymmetry $\alpha_{0}$ for $\Sigma^{+} \rightarrow p \pi^{0}$ is not
consistent with the $\Delta I = 1/2$ rule~\cite{Salone:2022lpt}, where
$\Delta I$ refers to the isospin difference between the initial and
final states.  Therefore, a precision measurement of the decay
asymmetry $\alpha_+$ for $\Sigma^{+} \rightarrow n \pi^{+}$ is needed
to determine the contributions of the $\Delta I=3/2$ and $\Delta
I=3/2$ weak transitions to $\Sigma$ decays~\cite{Overseth:1969bxc}.

Experimentally, the decay asymmetry parameters $\alpha_{0}$ and its
charge-conjugated (c.c.) equivalent $\bar{\alpha}_{0}$ have been well
measured~\cite{ParticleDataGroup:2020ssz}.  For the decay of
$\Sigma^{+} \rightarrow n \pi^{+}$, there are only two measurements
of $\alpha_+$ from fixed-target experiments, performed more than fifty years
ago. Although the two existing results $0.069 \pm
0.017$~\cite{osti_4750337} and $0.037 \pm 0.049$~\cite{Berley:1970zf}
are in agreement with each other, they are relatively imprecise
compared with $\alpha_{0}$ and also compatible with zero. Furthermore,
the corresponding decay parameter $\bar{\alpha}_{-}$ of
$\bar\Sigma^{-} \rightarrow \bar{n} \pi^{-}$ has never been measured
before.  Precision measurements of the decay parameters of $\Sigma^{+}
\rightarrow n \pi^{+}$ + c.c. would provide a first precision test of
$C\!P$ symmetry in hyperon to neutron decays and supply important
experimental input to sharpen $C\!P$ violation predictions of all
non-leptonic two-body weak decays of $\Sigma$. However, the relatively
small $\alpha_{+}$ and $\bar{\alpha}_{-}$ values, the $\Sigma$
polarization value determination, and the difficulties in neutron and
anti-neutron detection all represent a challenge for an accurate
experimental measurement.

\begin{figure}[htbp]
\begin{center}
\begin{overpic}[width=0.48\textwidth,angle=0]{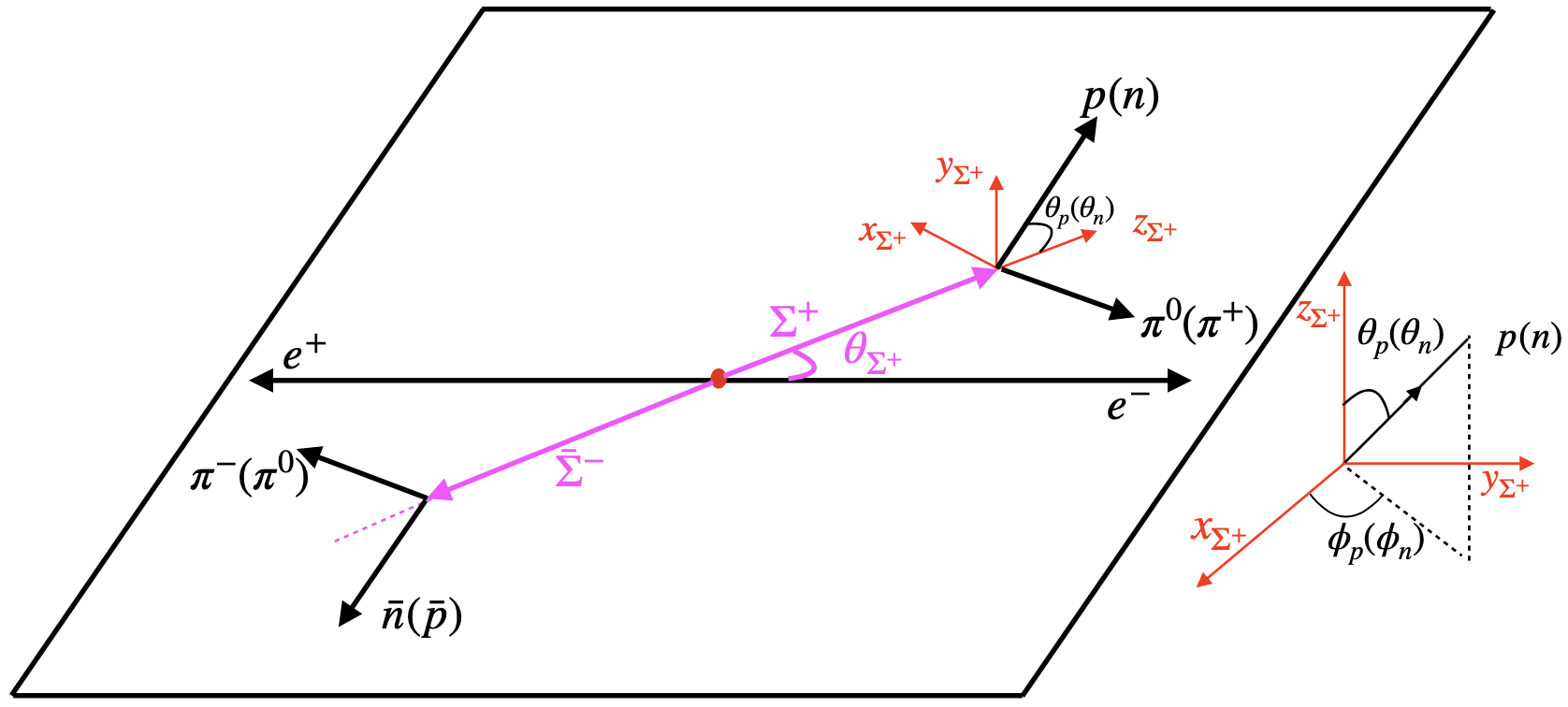}
\end{overpic}
\end{center}
\caption{The helicity frame definitions for $J/\psi \to\Sigma^+\bar\Sigma^-$,$\Sigma^+\to p \pi^0(n\pi^+)$,
 $\bar\Sigma^-\to \bar{n} \pi^-(\bar{p} \pi^0)$. In the $e^{+} e^{-}$ center-of-mass system, $\theta_{\Sigma}$ is the angle between the $\Sigma^{+}$ and the electron beam direction. The $z_{\Sigma^+}$ axis is the moving direction of $\Sigma^{+}$ in the $J/\psi$ rest frame, the $y_{\Sigma^+}$ axis is perpendicular to the plane of $\Sigma^{+}$ and electron, and the $x_{\Sigma^+}$ axis is defined by the right-handed coordinate system.}
\label{fig:helicity_frame}
\end{figure}

The BESIII experiment provides a unique environment to study both
hyperon production and decay properties in electron-positron
annihilation to $\Sigma^{+}$$\bar{\Sigma}^{-}$ pairs via the
intermediate $J/\psi$ resonance, where the above challenges can be
well addressed. In this entangled quantum system, the decay parameters
of $\Sigma^{+}$ and $\bar \Sigma^{-}$ are correlated, allowing for a
precise determination of the asymmetry parameters and
the $C\!P$ symmetry.  The $e^{+}e^{-}\rightarrow J/\psi \to
\Sigma^{+}\bar{\Sigma}^{-}$ process is described by the $\Psi$
electric and magnetic form factors, $G_{E}^{\Psi}$ and $G_{M}^{\Psi}$
~\cite{Faldt:2017kgy}. These two $\Psi$ form factors are formally
equivalent to the $\Sigma$ electric and magnetic form
factors~\cite{Dubnickova:1992ii,Gakh:2005hh,Czyz:2007wi,Faldt:2013gka,Faldt:2016qee}. They
are usually parameterized by two real parameters $\alpha_{J/\psi}$ and
$\Delta\Phi$, which correspond to the angular decay asymmetry and the
relative phase between the two form factors, respectively. The
observable $\Delta\Phi$ is related to the spin-polarization of the
produced $\Sigma^{+} \bar{\Sigma}^{-}$ pair. The $\Sigma$ polarization
is perpendicular to the production plane and depends on the opening
angle $\theta_{\Sigma^{+}}$ between the $\Sigma^{+}$ and electron
($e^{-}$) beam in the reaction center-of-mass frame, as shown in
Fig.~\ref{fig:helicity_frame}.  The first polarization measurement of
$J/\psi \to \Sigma^{+} \bar{\Sigma}^{-}$ was reported by the BESIII
collaboration with $\Sigma^{+} \rightarrow p \pi^{0}$ and
$\bar\Sigma^{-} \rightarrow \bar{p} \pi^{0}$ based on 1.3 billion
$J/\psi$ events~\cite{BESIII:2020fqg}. The significant polarization
provides the prerequisite for $\alpha_{+}$ and $\bar{\alpha}_{-}$
measurements.

The production and decay process $e^{+}e^{-}\rightarrow J/\psi
\rightarrow \Sigma^{+}(\to N \pi) \bar{\Sigma}^{-}(\to\bar{N}\pi )$ is
described with five observables $\boldsymbol{\xi}= (
\theta_{\Sigma^{+}}, \theta_{N}, \phi_{N}, \theta_{\bar{N}},
\phi_{\bar{N}})$~\cite{Faldt:2017kgy}.  Here $\theta_{N}, \phi_{N}$
and $\theta_{\bar{N}}, \phi_{\bar{N}}$ are the polar and azimuthal
angles of the nucleon and anti-nucleon measured in the rest frames of
their respective mother particles.  The differential cross section
${\cal{W}}({\boldsymbol{\xi}})$ is defined as

\begin{equation*}
\small
\begin{split}
	{\cal{W}}({\boldsymbol{\xi}})=	 &{\cal{T}}_0({\boldsymbol{\xi}})+{{\alpha_{J/\psi}}}{\cal{T}}_5({\boldsymbol{\xi}})\\
	+&{{\alpha}{\bar{\alpha}}}\left({\cal{T}}_1({\boldsymbol{\xi}})
+\sqrt{1-\alpha_{J/\psi}^2}\cos({{\Delta\Phi}}){\cal{T}}_2({\boldsymbol{\xi}})
+{{\alpha_{J/\psi}}}{\cal{T}}_6({\boldsymbol{\xi}})\right)\\
+&\sqrt{1-\alpha_{J/\psi}^2}\sin({{\Delta\Phi}})
\left({\alpha}{\cal{T}}_3({\boldsymbol{\xi}})
+\bar{{\alpha}}{\cal{T}}_4({\boldsymbol{\xi}})\right) ,\label{eq:anglW}
\end{split}
\end{equation*}
where ${\cal{T}}_{i}, (i = 0, 1...6)$ are angular functions dependent
on $\boldsymbol{\xi}$ and described in detail in
Ref.~\cite{Faldt:2017kgy}. 
According to the above cross-section formula, if the process
of $J/\psi \to\Sigma^+\bar\Sigma^-$ with $\Sigma^+\to n\pi^+$,
$\bar\Sigma^-\to \bar{n} \pi^-$ is used with the $\alpha$ and
$\bar{\alpha}$ parameters close to zero, the cross section will be
small, and the determination of the parameters imprecise.  Moreover,
the simultaneous detection of both the neutron and anti-neutron will
be difficult.  In addition, the process of $J/\psi
\to\Sigma^-\bar\Sigma^+$ with $\Sigma^-\to n\pi^-$, $\bar\Sigma^+\to
\bar{n} \pi^+$ with the same final state could contaminate our signal.
To overcome these disadvantages, we instead use
$J/\psi \rightarrow \Sigma^+\bar\Sigma^-$ with $\Sigma^{+}\to
p\pi^{0}$($n\pi^{+})$ and $\bar\Sigma^{-}\to
\bar{n}\pi^{-}$($\bar{p}\pi^{0})$.  Benefiting from the large decay
parameters $\alpha_{0}=-0.998\pm0.037_{\rm stat}\pm0.009_{\rm syst}$
and $\bar{\alpha}_{0}=0.990\pm0.037_{\rm stat}\pm0.011_{\rm
  syst}$~\cite{BESIII:2020fqg}, the measurement accuracy can be
improved by 17.4 times compared with the neutron anti-neutron final
state. Also, since $\Sigma^{-}$ cannot decay to $\bar{p} \pi^{0}$, the
$J/\psi \to\Sigma^-\bar\Sigma^+$ background is highly suppressed.

This Letter is based on a data sample of
$(1.0087\pm0.0044)\times10^{10}$ $J/\psi$ events~\cite{BESIII:2021cxx}
taken with the BESIII detector operating at the BEPCII
collider. Details about the design and performance of the BESIII
detector are given in Ref.~\cite{Ablikim:2009aa}.  Candidate events
for the process $J/\psi \rightarrow \Sigma^+\bar\Sigma^{-}$ with
subsequent $\Sigma^{+} \rightarrow p\pi^{0} (n\pi^{+})$ and
$\bar\Sigma^{-} \rightarrow \bar{n}\pi^{-} (\bar{p}\pi^{0})$ decays
must have two charged tracks with opposite charges and at least two
photons. Charged tracks detected in the multi-layer drift chamber
(MDC) are required to be within a polar angle ($\theta$) range of
$|\cos \theta|<0.93$, where $\theta$ is defined with respect to the
$z$ axis, the symmetry axis of the MDC.  For each track, the distance
of the closest approach to the interaction point must be less than 10
cm along the $z$ axis, and less than 2 cm in the transverse plane.

The particle identification (PID) system identifies the two candidate
charged tracks as $p \pi^{-}$ or $\bar{p} \pi^{+}$ based on the
measured energy loss in the MDC and the flight time in the
time-of-flight system.  Each track is assigned to the particle type
corresponding to the hypothesis with the highest confidence level.

Photon candidates are identified using showers in the electromagnetic
calorimeter (EMC). The deposited energy of each shower must be more
than 25 MeV in the barrel region ($ |\cos \theta| < 0.80$) and more
than 50 MeV in the end-cap region ($ 0.86 <|\cos \theta| <
0.92$). To exclude showers that originate from charged tracks, the
opening angle subtended by the EMC shower and the position of the
closest charged track at the EMC must be greater than $10^\circ$ as
measured from the interaction point.  To suppress electronic noise and
showers unrelated to the event, the difference between the EMC time
and the event start time is required to be within [0, 700] ns.

Candidates for $\pi^{0}$ are selected as photon pairs with an
invariant mass in the interval of $(m_{\pi^0}-60~\mevcc) <M_{\gamma
  \gamma} < (m_{\pi^0}+40~\mevcc)$, where $m_{\pi^0}$ is the known
$\pi^0$ mass~\cite{ParticleDataGroup:2020ssz}.
In addition, a one-constraint (1C) kinematic fit
is performed on the selected photon pairs, constraining the invariant
mass to the known $\pi^0$ mass. The $\chi^2_{\rm 1C}$ of the kinematic
fit is required to be less than 25.  At least one candidate $\pi^0$ is
required.

To select $J/\psi\to\Sigma^+\bar\Sigma^-$ with $\Sigma^+\to p\pi^0$
and $\bar\Sigma^-\to \bar n\pi^-$, the anti-neutron energy deposition
in the EMC is required to be at least 0.5 GeV. The second moment,
defined as $\sum_i E_i r_i^2 /\sum_i E_i$, is required to be greater
than 20. Here $E_i$ is the energy deposition in the $i_\mathrm{th}$
crystal and $r_i$ is the radial distance of the $i_\mathrm{th}$ crystal from the
cluster center. The opening angle $\theta_{\gamma,\bar n}$ between
photon candidates and the $\bar n$ track is required to be greater
than $20^{\circ}$.  For this process, a four-constraint (4C) kinematic
fit is applied by imposing energy-momentum conservation and an
additional $\pi^{0}$ mass constraint, where the direction of the $\bar
n$ is measured and the energy is unmeasured.  A two-constraint (2C)
kinematic fit is applied to the $J/\psi \to \Sigma^+\bar\Sigma^-$
process, with $\Sigma^+\to n\pi^+$ and $\bar\Sigma^-\to \bar
p\pi^0$. Energy-momentum conservation and an additional $\pi^{0}$ mass
constraint are imposed in this fit, with the neutron being treated as
a missing particle. The 4C and 2C kinematic fit chi-squares,
$\chi^2_{\rm 4C}$ and $\chi^2_{\rm
  2C}$, are both required to be less than 100. If the number of $\pi^0$
candidates in an event is more than one, the combination with the
minimum $\chi^{2}_{\rm 4C}$ or $\chi^{2}_{\rm 2C}$ is selected as the
final candidate.

\begin{figure*}[hbtp]
\subfigure{
\begin{minipage}{\textwidth}
\centering
\includegraphics[width=0.4\textwidth]{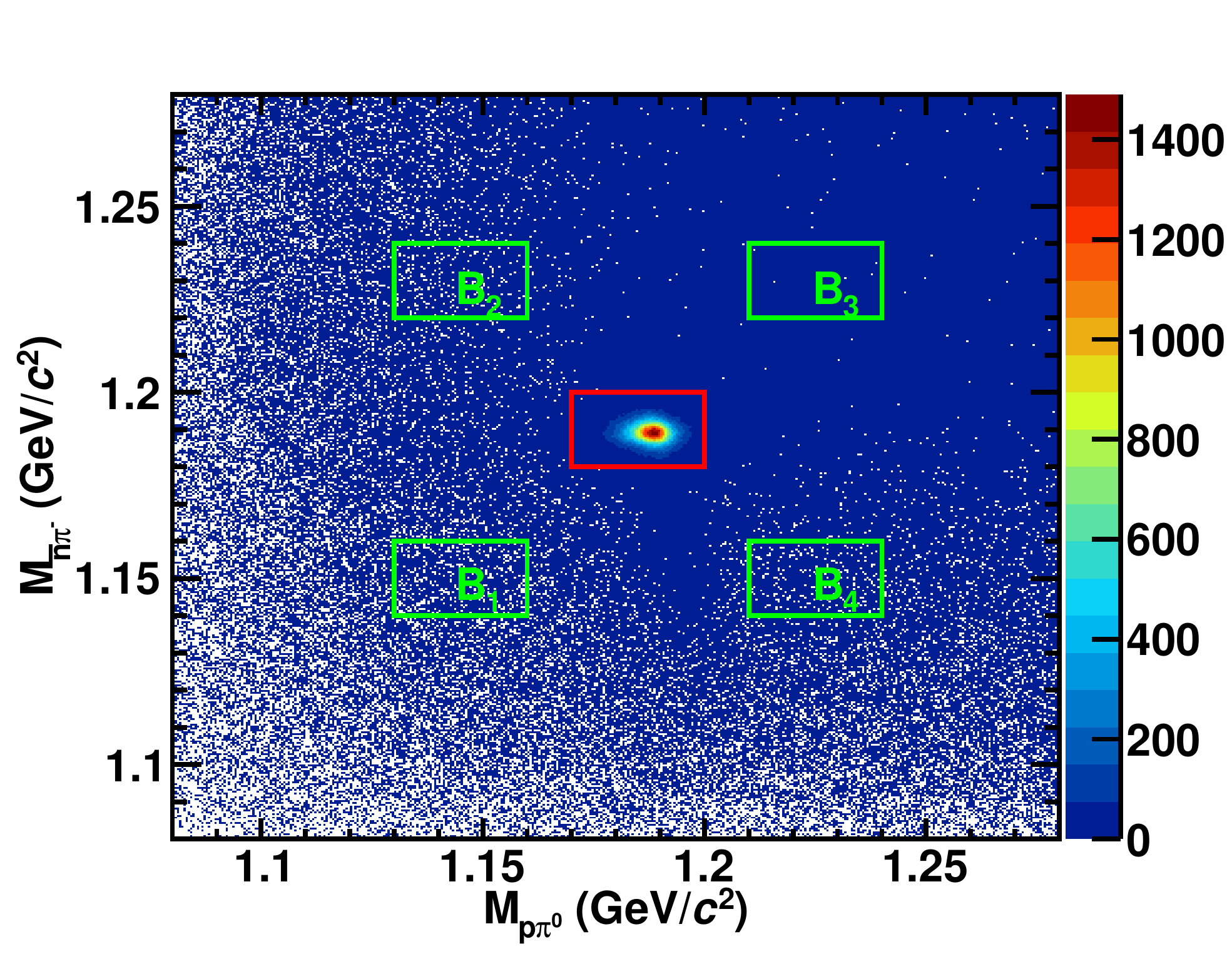}
\includegraphics[width=0.4\textwidth]{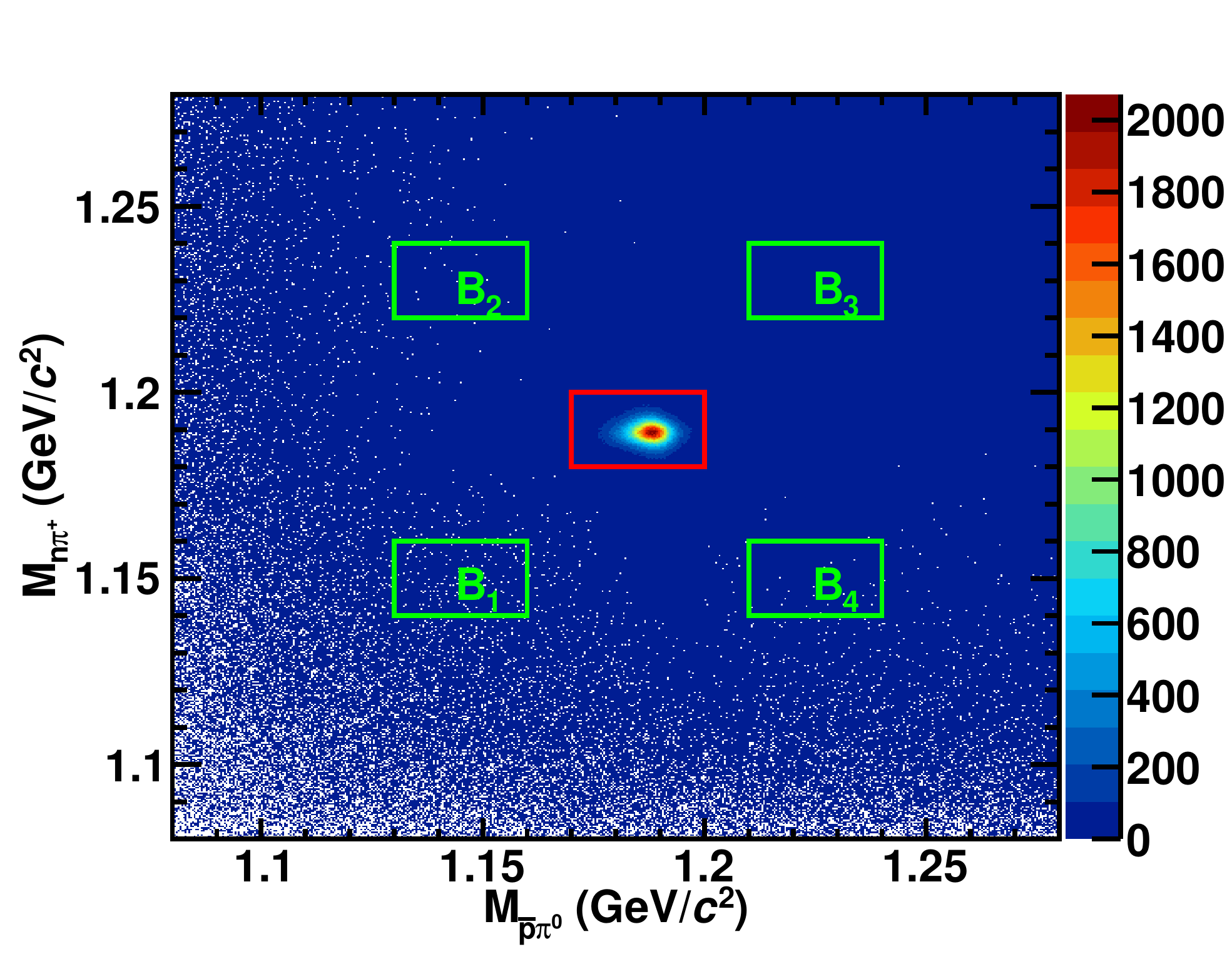}
\end{minipage}}
\caption{Distributions of (left) $M_{\bar n \pi^{-}}$ versus $M_{p \pi^{0}}$ for $J/\psi\to\Sigma^{+}\bar{\Sigma}^{-}$ with $\Sigma^{+} \rightarrow p \pi^{0}, \bar{\Sigma}^{-} \rightarrow \bar{n} \pi^{-}$, and (right) $M_{n \pi^{+}}$ versus $M_{\bar p \pi^{0}}$ for $J/\psi\to\Sigma^{+}\bar{\Sigma}^{-}$ with $\Sigma^{+} \rightarrow n \pi^{+}, \bar{\Sigma}^{-} \rightarrow \bar{p} \pi^{0}$. The red boxes denote the signal regions and the green ones indicate the sideband regions.}
\label{fig:scatter}
\end{figure*}

To investigate possible background after applying the event selection
criteria, an inclusive Monte Carlo (MC) sample of 10 billion $J/\psi$
events has been examined with TopoAna~\cite{ref:topo}. All particle
decays are modeled with {\sc evtgen}~\cite{ref:evtgen} using branching
fractions either taken from the Particle Data Group
(PDG)~\cite{ParticleDataGroup:2020ssz}, when available, or otherwise
estimated with {\sc lundcharm}~\cite{ref:lundcharm}.  The main peaking
backgrounds are $J/\psi\to\gamma\Sigma^+\bar\Sigma^-$ and
$J/\psi\to\gamma\eta_c, \eta_c\to\Sigma^+\bar\Sigma^-$, which both
contribute 0.2\% of the signal strength and are negligible.  The
non-peaking background mainly includes
$J/\psi\to\Delta^+\bar\Delta^-\to p\pi^0\bar n \pi^-(n\pi^+\bar
p\pi^0)$ and $J/\psi\to p\pi^0\bar n \pi^-(n\pi^+\bar p\pi^0)$ whose
contributions are estimated to be 1.4\% and 1.6\% with a two-dimensional
sideband method.  Figure~\ref{fig:scatter} shows the distributions of
$M_{\bar n \pi^{-}}$ versus $M_{p \pi^{0}}$ and $M_{n \pi^{+}}$ versus
$M_{\bar p \pi^{0}}$ for the two decay modes.  The signal regions in
the red rectangles are defined as: 1.17 $<M_{p\pi^0}(M_{\bar
  p\pi^0})<$ 1.20 GeV/$c^2$ and 1.18 $<M_{\bar n\pi^-}(M_{n\pi^+})<$
1.20 GeV/$c^2$.  To estimate the non-peaking background contributions,
four sideband regions have been selected, denoted as green rectangles
in the plots. Each sideband region has the same area as the signal
region and is placed at a distance of about $2\sigma$ from the signal
boundary, where $\sigma$ is the invariant mass resolution of
$\Sigma^+$ and $\bar\Sigma^-$. The background events are estimated
using $f\times \sum_{i=1}^{4}B_{i}$, where
$B_{i}$ is the number of events in the $i_{\rm th}$ sideband
region, and the scale factor $f$ is defined as the background ratio
between the signal and sideband regions. Using a two-dimensional fit on
the distribution of $M_{p\pi^0}$ versus $M_{\bar n\pi^-}$ or
$M_{n\pi^+}$ versus $M_{\bar p \pi^0}$, the scale factors are determined
to be $0.265\pm0.001$ and $0.259\pm0.001$ for these two decay
channels, respectively. Here the uncertainties are statistical only.

\begin{figure}[htbp]
\includegraphics[width=0.49\textwidth]{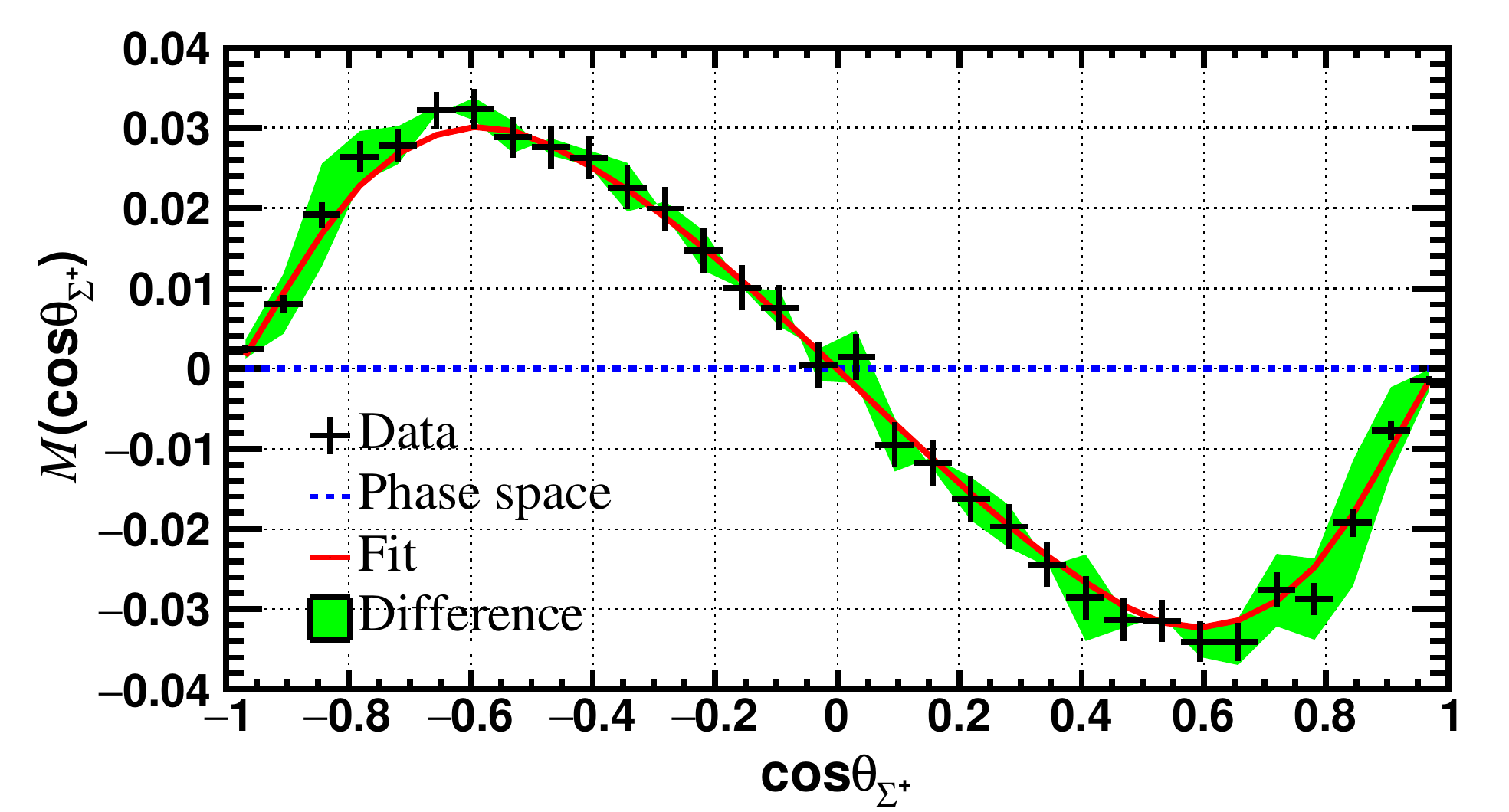}
\caption{The moment $M(\cos\theta_{\Sigma^{+}})$ for data, that is not
  corrected for acceptance and reconstruction efficiency, as a function
  of $\cos\theta_{\Sigma^{+}}$ for
  $J/\psi\to\Sigma^{+}\bar{\Sigma}^{-}$ with the two decay channels:
  $\Sigma^{+} \rightarrow p \pi^{0}, \bar{\Sigma}^{-} \rightarrow
  \bar{n} \pi^{-}$ and $\Sigma^{+} \rightarrow n \pi^{+},
  \bar{\Sigma}^{-} \rightarrow \bar{p} \pi^{0}$. The black points with
  error bars are data with background subtracted, the red solid line
  is the fit result and the blue dashed line represents the
  distribution without polarization uniformly distributed in phase
  space. The height of green band shows the absolute difference between the two decay
  channels with background subtracted.}
\label{fig:polarization}
\end{figure}

An unbinned maximum likelihood fit is performed in the five angular
dimensions ${\boldsymbol{\xi}}$~\cite{BESIII:2020fqg} simultaneously
on the two datasets to determine the parameters $\{\alpha_{J/\psi},
\Delta\Phi_{J/\psi}, \alpha_{+}, \bar{\alpha}_{-}\}$. Following the approach in Ref.~\cite{BESIII:2018cnd}, the
multidimensional fit takes the reconstruction efficiency into
account in a model-independent way and background contribution has been considered according to the scale factors $f$.  The numerical fit results are
summarized in Table~\ref{table:sum_decay}. The relative phase
between the $\Psi$ electric and magnetic form factors is determined to
be $\Delta\Phi_{J/\psi} = (-0.2772 \pm 0.0044_{\rm
  stat}\pm0.0041_{\rm syst})$ rad, which implies
$\Sigma$ spin polarization is observed. The moment related to the
polarization is defined as

\begin{equation*}
M(\cos\theta_{\Sigma^{+}}) = \frac{m}{n}\sum_i^{n_{k}}(\sin\theta_{N}^{k} \sin\phi_{N}^{k} - \sin\theta^{k}_{\bar{N}} \sin\phi_{\bar{N}}^{k}).
\end{equation*}
Here, $m=40$ is the number of bins, $n$ is the total number of events
in the data sample, and $n_{k}$ is the number of events in the $k_\mathrm{th}$ 
$\cos\theta_{\Sigma^{+}}$ bin.  The expected angular dependence of the
moment is
$\frac{dM}{d\cos\theta_{\Sigma^{+}}}\sim\sqrt{1-\alpha_{J/\psi}^2}\alpha_{+}
\sin\Delta\Phi_{J/\psi} \cos\theta_{\Sigma^{+}} \sin\theta_{\Sigma^{+}}$.  In
Fig.~\ref{fig:polarization}, the black points represent data and
follow the expectation as shown by the red line.  As $\Delta\Phi_{J/\psi}$ is
not zero, it is possible to determine the asymmetry parameters
$\alpha_{+}$ and $\bar{\alpha}_{-}$ simultaneously.  The asymmetry
decay parameter $\alpha_{+}$ is measured to be $0.0481\pm0.0031_{\rm
  stat}\pm0.0019_{\rm syst}$, with a precision improved by a factor of
4.7 compared to the previous best measurement~\cite{osti_4750337}.
The asymmetry decay parameter $\bar{\alpha}_{-}$ is determined for the
first time as $-0.0565\pm0.0047_{\rm stat}\pm0.0022_{\rm syst}$.
Assuming no $C\!P$ violation, the average decay asymmetry is
calculated to be $\left< \alpha_{+}\right>=(\alpha_{+}-
\bar\alpha_{-})/2 = 0.0506\pm0.0026_{\rm stat}\pm0.0019_{\rm syst}$.

\begin{table}[hbtp]
 \centering
 \footnotesize
\caption{The decay parameters of $J/\psi\rightarrow\Sigma^+\bar\Sigma^-,\Sigma^+\rightarrow p\pi^0(n\pi^+), \bar\Sigma^-\rightarrow\bar n \pi^-(\bar p \pi^0)$. The first uncertainties are statistical and the second systematic. Dash ($-$) represents no experimental measurement.} 
\label{table:sum_decay}
\begin{tabular}{lccc}
\hline \hline 
Parameter&This work&Previous result\\\hline 
$\alpha_{J/\psi}$&$-0.5156\pm0.0030\pm0.0061$&$-0.508\pm0.006\pm0.004$~\cite{BESIII:2020fqg}\\ 
$\Delta\Phi_{J/\psi}$(rad)&$-0.2772\pm0.0044\pm0.0041$&$-0.270\pm0.012\pm0.009$~\cite{BESIII:2020fqg}\\
$\alpha_{+}$&$\phantom{-}0.0481\pm0.0031\pm0.0019 $&$0.069\pm0.017$~\cite{osti_4750337} \\
$\bar{\alpha}_{-}$&$-0.0565\pm0.0047\pm0.0022 $&$-$ \\\hline 
$\alpha_{+}/\alpha_{0}$&$-0.0490\pm0.0032\pm0.0021 $&$-0.069\pm0.021 $~\cite{Marraffino:1980dj} \\
$\bar{\alpha}_{-}/\bar\alpha_{0}$&$-0.0571\pm0.0053\pm0.0032 $&$-$\\ 
$\it{A_{C\!P}}$&$-0.080\pm0.052\pm0.028$&$-$\\
$\it{\left< \alpha_{+}\right>}$&$\phantom{-}0.0506\pm0.0026\pm0.0019$&$-$ \\
 \hline \hline
\end{tabular}
\end{table}

 \begin{table}[hbtp]
 \centering
\caption{The absolute systematic uncertainties in $\alpha_{J/\psi}$, $\Delta\Phi_{J/\psi}$, $\bar{\alpha}_{-}$ and $\alpha_{+}$.} 
\label{tot_parameter1}
\begin{tabular}{lccccc}
\hline \hline 
Source &$\alpha_{J/\psi}$ &$\Delta\Phi_{J/\psi}$&$\bar{\alpha}_{-}$&$\alpha_{+}$\\\hline 
MC efficiency correction&$0.0059$ &$0.0005 $&$0.0016$&$0.0011$ \\ 
Kinematic fit&$0.0003$ &$0.0004 $&$0.0007$&$0.0003$ \\ 
Signal mass window&$0.0015$ &$0.0021 $&$0.0010$&$0.0009$\\
Background&$0.0001$ &$0.0007$&$0.0003$&$0.0002$\\\hline
Fitting method&$0.0007 $ &$0.0028$&$0.0007$&$0.0012$\\
Decay parameters&$0.0000$ &$0.0020$&$0.0003$&$0.0003$\\\hline
 Total&$0.0061$ & $0.0041$  &$0.0022$&$0.0019$\\
 \hline \hline
\end{tabular}
\end{table}

The systematic uncertainties are listed in Table~\ref{tot_parameter1},
which is divided into two categories. The first category is from the
event selection, including the uncertainties for MDC tracking, PID,
$\pi^{0}$ and $\bar{n}$ reconstructions, kinematic fit, background
estimations, as well as the $\Sigma^{+}$ and $\bar{\Sigma}^{-}$ mass
window requirements. The second category includes the uncertainties
associated with the fit procedure. The individual uncertainties are
assumed to be uncorrelated and are therefore added in quadrature.  The
uncertainties due to potential efficiency differences between data and
simulation for charged-particle tracking and PID have been
investigated with a $J/\psi\to p\bar p\pi^+\pi^-$ control sample, and
those due to neutral $\pi^{0}$ and $\bar{n}$ reconstructions are
estimated from $J/\psi\to \Sigma^+(p\pi^0) \bar\Sigma^-(\bar p\pi^0)$
and $J/\psi\to p\bar n\pi^-$ control samples. Using these control
samples, we determine the correction factors and apply them in the MC
simulation to obtain the nominal results.  The uncertainty of the
correction factors is estimated by changing them within $1\sigma$
regions. The differences to the nominal results are taken as the MC
efficiency correction systematic uncertainties.  The systematic
uncertainties due to the kinematic fits are examined by comparing the
detection efficiencies with and without helix parameter corrections,
which are used to reduce the discrepancies between data and MC
simulation~\cite{BESIII:2012mpj}. The differences in detection
efficiencies with and without corrections are assigned as the
systematic uncertainties.  To estimate the systematic uncertainty
associated with the signal mass window, the window is changed by
$3\sigma$ ($\pm 5~\mev$), where $\sigma$ is the invariant mass
resolution of $\Sigma^+$ and $\bar\Sigma^-$. The fits are repeated
using the new mass window, and the differences of results to the
nominal values are regarded as the corresponding systematic
uncertainties. The systematic uncertainty caused by the background
estimation is studied by varying the length and width of four sideband
boxes within $\pm 5~\mev$. The largest differences in the parameters
are taken as the systematic uncertainties.  To validate the
reliability of the fit results, a set of 100 pseudo-data samples are
simulated and subjected to the same selection criteria. In these
samples, the differential cross section is based on the decay
parameters listed in Table \ref{table:sum_decay}. The systematic
uncertainties from the fit approach are assumed to be the deviations
between the inputs and average outputs.  In the nominal fit, the
parameters of $\alpha_{0}$ and $\bar\alpha_{0}$ are fixed at the world
average values~\cite{ParticleDataGroup:2020ssz}. By changing the
parameter within $\pm 1\sigma$ (0.006) regions for $\alpha_{0}$ and
$\bar\alpha_{0}$, the 
changes between the new and nominal fit
results are taken as systematic uncertainties.

In summary, based on a data sample of $(1.0087\pm0.0044)\times10^{10}$
$J/\psi$ events collected at the BESIII detector, the five-dimensional
angular analysis of the processes of $J/\psi \to\Sigma^+\bar\Sigma^-$
($\Sigma^+\to p \pi^0, \bar\Sigma^-\to \bar n \pi^- $ and $\Sigma^+\to
n \pi^+, \bar\Sigma^-\to \bar p \pi^0 $) is performed. The decay
parameters $\alpha_{J/\psi}$ and $\Delta\Phi_{J/\psi}$ are measured to
be $-0.5156\pm0.0030_{\rm stat}\pm0.0061_{\rm syst}$ and
$(-0.2772\pm0.0044_{\rm stat}\pm0.0041_{\rm syst})$ rad, respectively,
which are consistent with the previous measurements but with improved
precision~\cite{BESIII:2020fqg}. The nonzero value of
$\Delta\Phi_{J/\psi}$ in the $J/\psi \to\Sigma^+\bar\Sigma^-$ decay,
which implies the existence of polarization, is confirmed with two
different $\Sigma$ decay channels, $J/\psi \to \Sigma^+\bar\Sigma^-
\to p \pi^0\bar{n}\pi^-(n\pi^+\bar{p}\pi^0)$ and $J/\psi \to
\Sigma^+\bar\Sigma^- \to p \pi^0\bar{p}\pi^0$. Therefore, the decay
asymmetry parameters $\alpha_{+}$ and $\bar\alpha_{-}$ are determined
simultaneously. The parameters $\alpha_{+}$ and
$\alpha_{+}/\alpha_{0}$ determined in this work are consistent with
the PDG averages but with significantly improved precision, and
$\bar{\alpha}_{-}$ and $\bar{\alpha}_{-}/\bar{\alpha}_{0}$ are
measured for the first time. The average decay asymmetry parameter is
$0.0506 \pm 0.0026_{\rm stat}\pm 0.0019_{\rm syst}$, which differs from
zero by $16\sigma$. In contrast, the previous best measurement only
deviates from zero by about $3\sigma$~\cite{osti_4750337}. This result is
crucial to test the $\left | \Delta I \right | = 1/2$ rule and study
the high order isospin transitions~\cite{Marraffino:1980dj}. Our
precise measurement of the decay asymmetry parameter in the neutron
mode is of vital importance to the $C\!P$ violation
prediction~\cite{Tandean:2002vy}.  This is the first study to test
$C\!P$ symmetry in the hyperon to neutron decay, and the result is
consistent with $C\!P$-conservation.

The BESIII Collaboration thanks the staff of BEPCII and the IHEP computing center for their strong support. This work is supported in part by National Key R\&D Program of China under Contracts Nos. 2020YFA0406300, 2020YFA0406400; National Natural Science Foundation of China (NSFC) under Contracts Nos. 11635010, 11735014, 11835012, 11935015, 11935016, 11935018, 11961141012, 12022510, 12025502, 12035009, 12035013, 12061131003, 12192260, 12192261, 12192262, 12192263, 12192264, 12192265, 12221005, 12225509, 12235017; the Chinese Academy of Sciences (CAS) Large-Scale Scientific Facility Program; the CAS Center for Excellence in Particle Physics (CCEPP); CAS Key Research Program of Frontier Sciences under Contracts Nos. QYZDJ-SSW-SLH003, QYZDJ-SSW-SLH040; 100 Talents Program of CAS; The Institute of Nuclear and Particle Physics (INPAC) and Shanghai Key Laboratory for Particle Physics and Cosmology; Sponsored by Shanghai Pujiang Program(20PJ1401700); ERC under Contract No. 758462; European Union's Horizon 2020 research and innovation programme under Marie Sklodowska-Curie grant agreement under Contract No. 894790; German Research Foundation DFG under Contracts Nos. 443159800, 455635585, Collaborative Research Center CRC 1044, FOR5327, GRK 2149; Istituto Nazionale di Fisica Nucleare, Italy; Ministry of Development of Turkey under Contract No. DPT2006K-120470; National Research Foundation of Korea under Contract No. NRF-2022R1A2C1092335; National Science and Technology fund of Mongolia; National Science Research and Innovation Fund (NSRF) via the Program Management Unit for Human Resources \& Institutional Development, Research and Innovation of Thailand under Contract No. B16F640076; Polish National Science Centre under Contract No. 2019/35/O/ST2/02907; The Swedish Research Council; the Olle Engkvist Foundation under contract no. 200-0605; U. S. Department of Energy under Contract No. DE-FG02-05ER41374.

\end{document}